\documentclass[preprint,12pt]{elsarticle}

\usepackage{graphicx}
\usepackage{epsfig}

\usepackage{amssymb}
\usepackage{amsthm}
\usepackage{amsmath}

\newcommand{\as}{\alpha_s}
\def\eq#1{{Eq.~(\ref{#1})}}
\def\fig#1{{Fig.~\ref{#1}}}

\journal{Nuclear Physics A}

\begin{document}

\begin{frontmatter}


  
  \title{Introduction to the Physics of Saturation}

  
  \author{Yuri V. Kovchegov\footnote{This review is based on the talk
      given by the author at the RIKEN BNL Research Center Workshop on
      {\it Saturation, the Color Glass Condensate and Glasma: What
        Have we Learned from RHIC?} on May 10, 2010.}}

\address{Department of Physics,
  The Ohio State University,
  Columbus, OH 43210, USA}

\begin{abstract}
  We present a brief introduction to the physics of parton
  saturation/Color Glass Condensate (CGC). 
\end{abstract}


\end{frontmatter}



\section{Introduction}
\label{Intro}

One of the most interesting features of quantum chromodynamics (QCD)
is the property of the asymptotic freedom: the strong coupling
constant is small at large momenta/short distances, and it is large at
small momenta/large distances \cite{Gross:1973id,Politzer:1973fx}.
Finding the scale that determines the value of the characteristic
running QCD coupling is one of the the central questions for high
energy scattering physics, important for the theoretical description
of both the hadronic and nuclear scattering processes.

A naive answer to this question would be to say that in high energy
scattering the large center-of-mass energy $s$ determines the scale of
strong coupling making it small: $\as (s) \ll 1$. While such statement
would be true for several $s$-channel processes, the dominant
contribution to total cross sections in high energy hadronic and
nuclear scattering comes from the $t$-channel exchanges, for which the
scale of the QCD coupling constant is not given by $s$, but by the
typical transverse momentum in the problem. For proton-proton
scattering one may estimate the typical transverse momentum to be of
the order of the inverse transverse size of the protons, which is
roughly the QCD confinement scale $\Lambda_{QCD}$. People performing
such estimate would pessimistically conclude that the QCD coupling in
high energy hadronic scattering runs as $\as (\Lambda_{QCD}^2)$ and is
therefore not small, $\as (\Lambda_{QCD}^2) \sim 1$. With the coupling
constant of order--one, we would have little chance of describing the
total hadronic and nuclear cross sections from first principles, at
least certainly not with the help of QCD perturbation theory.

Traditional perturbative QCD (pQCD) approaches are well-aware of the
above problem, and try to avoid it by separating hard sub-events where
the coupling is small from the full event with large QCD coupling. For
instance, in hadronic scattering pQCD may be used to calculate jet
production cross section, where the hard partonic scattering is
factorized from the non-perturbative distribution and fragmentation
functions. The high transverse momentum $p_T$ of the produced hard
parton insures applicability of pQCD: $\as (p_T^2) \ll 1$.  Similarly,
in deep inelastic scattering (DIS) pQCD can describe structure
functions at high photon virtuality $Q^2$, since there $\as (Q^2) \ll
1$, but pQCD is expected to fail at low-$Q^2$. Since jet production
events in hadronic collisions constitute a small percentage of the
total cross section, pQCD approach describes only rare events, and is
not applicable to the description of the bulk particle production and
dynamics.

Saturation physics provides a new way of tackling the problem of total
hadronic and nuclear cross sections. It is based on the theoretical
observation that small-$x$ hadronic and nuclear wave functions, and,
therefore, the scattering cross sections as well, are described by an
internal momentum scale known as the {\sl saturation scale} and
denoted by $Q_s$ \cite{Gribov:1984tu}. This intrinsic momentum scale
grows with the center-of-mass energy $s$ in the problem, and with the
increasing atomic number of a nucleus $A$ (in the case of a nuclear
wave function) approximately as
\begin{align}\label{Qs}
  Q_s^2 \, \sim \, A^{1/3} \, s^\lambda
\end{align}
where the best current theoretical estimates of $\lambda$ give
$\lambda = 0.2 \div 0.3$ \cite{Albacete:2007sm}. Therefore, for
hadronic collisions at high energy and/or for collisions of large
ultrarelativistic nuclei, saturation scale becomes large, $Q_s^2 \gg
\Lambda^2_{QCD}$. Since for total cross sections $Q_s$ is usually the
only momentum scale in the problem, we expect it to give the scale of
the running QCD coupling constant, making it small
\begin{align}
  \as (Q_s^2) \, \ll \, 1
\end{align}
and allowing for first-principles calculations of total hadronic and
nuclear cross sections, along with extending our ability to calculate
particle production and to describe diffraction in a small-coupling
framework.

Below we present a short review of the saturation physics, which is
also known as the Color Glass Condensate (CGC) physics. For more
extensive descriptions of the subject we refer the readers to the
review articles
\cite{Jalilian-Marian:2005jf,Weigert:2005us,Iancu:2003xm}.


\section{Brief Review of Saturation Physics}

Traditional approach to saturation physics consists of two stages,
corresponding to two different levels of approximation.  The first
approximation corresponds to the classical gluon field description of
nuclear wave functions and scattering cross sections. It resums all
multiple rescatterings in the nucleus, but lacks rapidity-dependence.
The latter is included through quantum corrections, which are resummed
by the non-linear evolution equations. This constitutes the second
level of approximation. We will present both stages below.

\subsection{Classical Gluon Fields}

\begin{figure}[ht]
  \begin{center}
    \epsfig{file=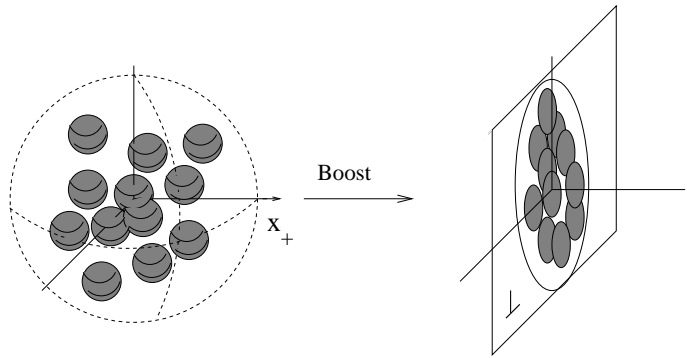, width=9cm} 
  \end{center}
\caption{Large nucleus before and after an ultrarelativistic boost.}
\label{nucl_boost}
\end{figure}

Imagine a single large nucleus, which was boosted to some
ultrarelativistic velocity, as shown in \fig{nucl_boost}. We are
interested in the dynamics of small-$x$ gluons in the wave function of
this relativistic nucleus. The small-$x$ gluons interact with the
whole nucleus coherently in the longitudinal direction: therefore,
only the transverse plane distribution of nucleons is important for
the small-$x$ wave function. As one can see from \fig{nucl_boost},
after the boost the nucleons, as ``seen'' by the small-$x$ gluons,
appear to overlap with each other in the transverse plane, leading to
high parton density. Large occupation number of color charges
(partons) leads to classical gluon field dominating the small-$x$ wave
function of the nucleus. This is the essence of the
McLerran-Venugopalan (MV) model \cite{McLerran:1993ni}. According to
the MV model, the dominant gluon field is given by the solution of the
classical Yang-Mills equations
\begin{align}\label{clYM}
  {\cal D}_\mu \, F^{\mu\nu} \, = \, J^\nu
\end{align}
where the classical color current $J^\nu$ is generated by the valence
quarks in the nucleons of the nucleus from \fig{nucl_boost}. 

\begin{figure}[ht]
  \begin{center}
    \epsfig{file=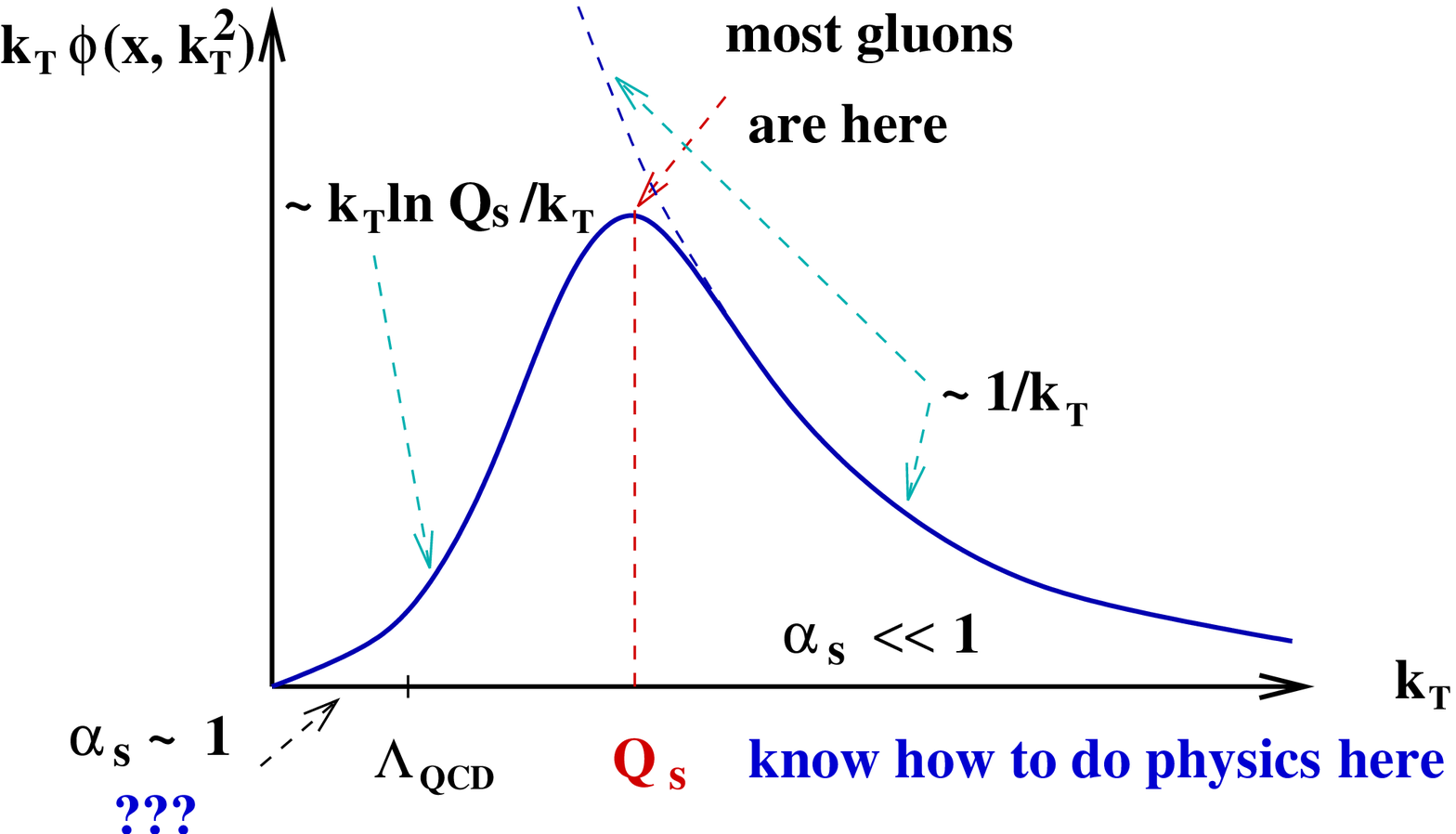, width=9.35cm} 
  \end{center}
\caption{Unintegrated gluon distribution $\phi (x, k_T^2)$ of a 
  large nucleus due to classical gluon fields (solid line). Dashed
  curve denotes the lowest-order perturbative result.}
\label{mv2}
\end{figure}

The equations (\ref{clYM}) were solved for a single nucleus exactly
\cite{Kovchegov:1996ty,Jalilian-Marian:1997xn} resulting in the
unintegrated gluon distribution $\phi (x, k_T^2)$ (multiplied by the
phase space factor of the gluon's transverse momentum $k_T$) shown in
\fig{mv2} as a function of $k_T$. (Note that in the MV model $\phi (x,
k_T^2)$ is independent of Bjorken-$x$.)
\fig{mv2} demonstrates the emergence of the saturation scale $Q_s$: as
one can see from \fig{mv2}, the majority of gluons in this classical
distribution have transverse momentum $k_T \approx Q_s$. Since in this
classical approximation $Q_s^2 \sim A^{1/3}$, for large enough nucleus
all of its small-$x$ gluons would have large transverse momenta $k_T
\approx Q_s \gg \Lambda_{QCD}$, justifying applicability of
perturbative approach to the problem. Note that the gluon distribution
slows down its growth with decreasing $k_T$ for $k_T < Q_s$ (from
power-law of $k_T$ to a logarithm): the distribution {\sl saturates},
justifying the name of the saturation scale.

\subsection{Nonlinear Evolution}

While the classical gluon fields of the MV model exhibit many correct
qualitative features of saturation physics, and give predictions about
$A$-dependence of observables which may be compared to the data, they
do not lead to any rapidity/Bjorken-$x$ dependence of the
corresponding observables, which is essential in the data on nuclear
and hadronic collisions. To include rapidity dependence one has to
calculate quantum corrections to the classical fields described above.

\begin{figure}[ht]
  \begin{center}
    \epsfig{file=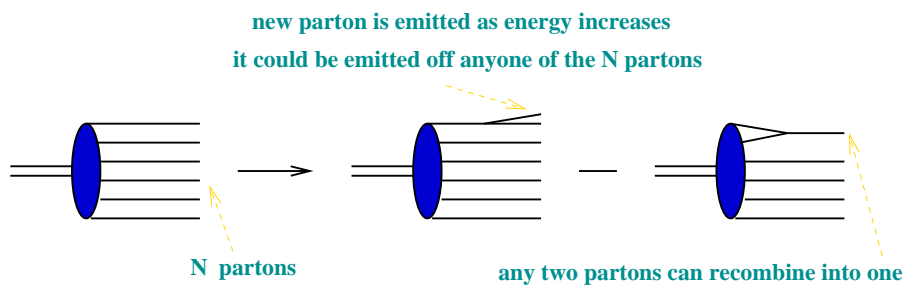, width=12cm} 
  \end{center}
\caption{Nonlinear small-$x$ evolution of a hadronic or nuclear wave 
  functions. All partons (quarks and gluons) are denoted by straight
  solid lines for simplicity.}
\label{BKfig}
\end{figure}

The inclusion of quantum corrections is accomplished by the small-$x$
evolution equations. The first small-$x$ evolution equation was
constructed before the birth of the saturation physics. This is the
Balitsky-Fadin-Kuraev-Lipatov (BFKL) evolution equation
\cite{Bal-Lip,Kuraev:1977fs}. This is a linear evolution equation,
which is illustrated by the first term on the right hand side of
\fig{BKfig}. Consider a wave function of a high-energy nucleus or
hadrons: it contains many partons, as shown on the left of
\fig{BKfig}. As we make one step of evolution by boosting the
nucleus/hadron to higher energy, either one of the partons can split
into two partons, leading to an increase in the number of partons
proportional to the number of partons $N$ at the previous step,
\begin{align}\label{BFKL}
  \frac{\partial \, N (x, k_T^2)}{\partial \ln (1/x)} \, = \, \as \,
  K_{BFKL} \, \otimes \, N (x, k_T^2),
\end{align}
with $K_{BFKL}$ an integral kernel. Clearly the BFKL equation
(\ref{BFKL}) introduces Bjorken-$x$/rapidity dependence in the
observables it describes.

The main problem with the BFKL evolution is that it leads to the
power-law growth of the total cross sections with energy,
$\sigma_{tot} \sim s^{\alpha_P -1}$, with the BFKL pomeron intercept
$\alpha_P -1 = (4 \, \as \, N_c \, \ln 2) /\pi >0$.  Such power-law
cross section increase violates the Froissart bound, which states that
the total hadronic cross section can not grow faster than $\ln^2 s$ at
very high energies.  Moreover, power-law growth of cross sections with
with energy violates the black disk limit known from quantum
mechanics: high-energy total scattering cross section of a particle on
a sphere of radius $R$ is bounded by
\begin{align}\label{bd}
  \sigma_{tot} \, \le \, 2 \, \pi \, R^2.
\end{align}
(Note the factor of 2 which is due to quantum mechanics, this is not
simply a hard sphere from classical mechanics!)

We see that something has to modify \eq{BFKL} at high energy. The
modification is illustrated on the far right of \fig{BKfig}: at very
high energies partons may start to recombine with each other on top of
the splitting. The recombination of two partons into one is
proportional to the number of pairs of partons, which, in turn, scales
as $N^2$. We end up with the following non-linear evolution equation:
\begin{align}\label{BK}
  \frac{\partial \, N (x, k_T^2)}{\partial \ln (1/x)} \, = \, \as \,
  K_{BFKL} \, \otimes \, N (x, k_T^2) - \as \, [N (x, k_T^2)]^2.
\end{align}
This is the Balitsky-Kovchegov (BK) evolution equation
\cite{Balitsky:1996ub,Kovchegov:1999yj}, which is valid for QCD in the
limit of large number of colors $N_c$. An equation of this type was
originally suggested by Gribov, Levin and Ryskin in
\cite{Gribov:1984tu} and by Mueller and Qiu in \cite{Mueller:1986wy},
though at the time it was assumed that the quadratic term is only the
first non-linear correction with higher order terms possibly appearing
as well: in \cite{Balitsky:1996ub,Kovchegov:1999yj} the exact form of
the equation was found, and it was shown that in the large-$N_c$ limit
\eq{BK} does not have any higher-order terms in $N$. Generalization of
\eq{BK} beyond the large-$N_c$ limit is accomplished by the
Jalilian-Marian--Iancu--McLerran--Weigert--Leonidov--Kovner (JIMWLK)
\cite{Jalilian-Marian:1997gr, Iancu:2000hn} evolution equation, which
is a functional differential equation. 

The physical impact of the quadratic term on the right of \eq{BK} is
clear: it slows down the small-$x$ evolution, leading to {\sl parton
  saturation} and to total cross sections adhering to the black disk
limit of \eq{bd}. The effect of gluon mergers becomes important when
the quadratic term in \eq{BK} becomes comparable to the linear term on
the right-hand-side. This gives rise to the saturation scale $Q_s$,
which now grows with energy (on top of its increase with $A$), as was
advertised in \eq{Qs} above.

\begin{figure}[ht]
  \begin{center}
    \epsfig{file=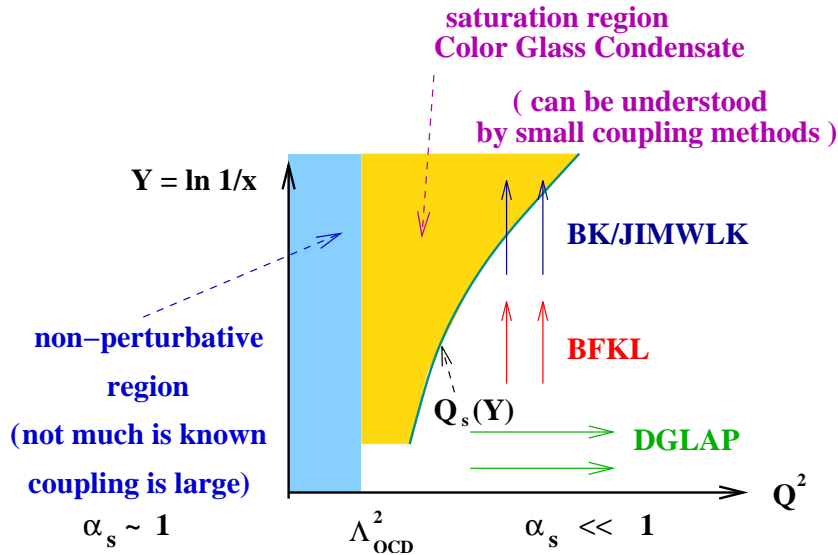, width=11cm} 
  \end{center}
\caption{Map of high energy QCD in the $(Q^2, Y=\ln 1/x)$ plane.}
\label{satbk}
\end{figure}

We summarize our knowledge of high energy QCD in \fig{satbk}, in which
different regimes are plotted in the $(Q^2, Y=\ln 1/x)$ plane, by
analogy with DIS. For hadronic and nuclear collisions one can think of
typical transverse momentum $p_T^2$ of the produced particles instead
of $Q^2$. Also rapidity $Y$ and Bjorken-$x$ variable are
interchangeable.  On the left of \fig{satbk} we see the region with
$Q^2 \le \Lambda_{QCD}^2$ in which the coupling is large, $\as \sim
1$, and small-coupling approaches do not work. In the pessimistic view
of high energy scattering described in the Introduction, this is
exactly where the total hadronic and nuclear cross sections would be.
In the perturbative region, $Q^2 \gg \Lambda_{QCD}^2$, we see the
standard DGLAP evolution and the linear BFKL evolution. The BFKL
equation evolves gluon distribution toward small-$x$, where parton
density becomes large and parton saturation sets in. Transition to
saturation is described by the non-linear BK and JIMWLK evolution
equations. Most importantly this transition happens at $Q_s^2 \gg
\Lambda_{QCD}^2$ where the small-coupling approach is valid.

\subsection{Some CGC Phenomenology}

One of the important predictions of saturation/CGC physics was the
so-called geometric scaling of the total DIS cross section. It was
argued that with $Q_s (x)$ being the only scale in the problem, DIS
structure functions and cross sections should depend only on one
variable -- $Q^2/Q_s^2 (x)$, instead of being functions of two
variables $x$ and $Q^2$. This prediction was supported by detailed
calculations based on BK evolution \cite{Levin:1999mw,Iancu:2002tr}
and was confirmed by an analysis of the HERA DIS data
\cite{Stasto:2000er}.

\begin{figure}[ht]
  \begin{center}
    \epsfig{file=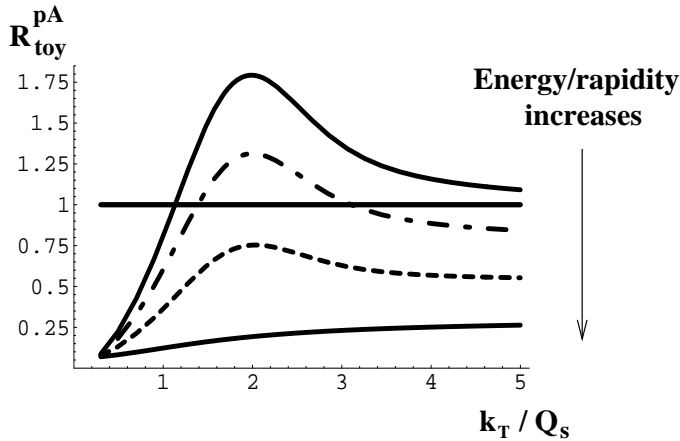, width=10cm} 
  \end{center}
\caption{Nuclear modification factor as a function of $k_T/Q_s$ as 
  predicted by saturation/CGC physics. Different curves correspond to
  different rapidities, with lower curves corresponding to higher
  rapidity.}
\label{toy}
\end{figure}

Another prediction of the non-linear evolution (\ref{BK}) concerns
particle production in proton-nucleus ($d+Au$) collisions. It follows
from \eq{BK} combined with the formulas for particle production in CGC
that the nuclear modification factor $R^{pA}$ should decrease as one
goes toward more forward rapidity
\cite{Kharzeev:2002pc,Kharzeev:2003wz,Albacete:2003iq}. This
prediction is illustrated in \fig{toy} and was confirmed by RHIC
experiments.


\section{Recent Progress in Saturation Physics}

In recent years saturation physics has become more precision-oriented,
with steps taken to improve its quantitative predictions. Running
coupling corrections to the BFKL/BK/JIMWLK evolution have been
calculated in \cite{Kovchegov:2006vj,Balitsky:2006wa}. It led to an
interesting result, in which the fixed coupling $\as$ in \eq{BK} was
replaced by a ``triumvirate'' of the running couplings at different
scales \cite{Kovchegov:2006vj,Balitsky:2006wa}:
\begin{align}
  \alpha_\mu \, \Longrightarrow \, \frac{\as (\ldots) \, \as
    (\ldots)}{\as (\ldots)}.
\end{align}
Such behavior has never been seen in field theories outside small-$x$
physics. Among other recent developments, next-to-leading-order BK
equation has been found in \cite{Balitsky:2008zz}, resulting in a
rather complicated but useful expression.

There is hope that we have finally managed to significantly advance
both the qualitative and quantitative understanding of QCD at high
energies.


\section*{Acknowledgments}

This work is supported in part by the U.S. Department of Energy under
Grant No. DE-FG02-10ER41687. 

The author would like to thank Anastasios Taliotis for providing him
with \fig{nucl_boost}.












\end{document}